\documentclass[12pt]{iopart}
\usepackage{txfonts}
\usepackage{graphicx}
\usepackage{amssymb}
\begin{document}

\title[]
{Growth and characterization of A$_{1-x}$K$_x$Fe$_2$As$_2$ (A = Ba,
Sr) single crystals with $x$=0 $\sim$ 0.4}

\author{Huiqian Luo, Zhaosheng Wang, Huan Yang, Peng Cheng, Xiyu Zhu and Hai-Hu Wen}

\ead{hhwen@aphy.iphy.ac.cn}

\address{
National Laboratory for Superconductivity, Institute of Physics and\\
National Laboratory for Condensed Matter Physics, P. O. Box 603
Beijing, 100190, P. R. China }

\begin{abstract}

Single crystals of A$_{1-x}$K$_x$Fe$_2$As$_2$ (A=Ba, Sr) with high
quality have been grown successfully by FeAs self-flux method.  The
samples have sizes up to 4 mm with flat and shiny surfaces. The
X-ray diffraction patterns suggest that they have high crystalline
quality and $c$-axis orientation. The non-superconducting crystals
show a spin-density-wave (SDW) instability at about 173 K and 135 K
for Sr-based and Ba-based compound, respectively. After doping K as
the hole dopant into the BaFe$_2$As$_2$ system, the SDW transition
is smeared, and superconducting samples with the compound of
Ba$_{1-x}$K$_x$Fe$_2$As$_2$ (0 $< x \leqslant$ 0.4) are obtained.
The superconductors characterized by AC susceptibility and
resistivity measurements exhibit very sharp superconducting
transition at about 36 K, 32 K, 27 K and 23 K, respectively.

\end{abstract}

\section{Introduction}
The newly discovered superconductivity in iron-oxypnictides
superconductors stimulates intensive researches on high-temperature
superconductivity beside the cuprate system. Just in several months,
the superconducting transition temperature ($T_c$) was promoted to
55 K in the electron doped system
\cite{Hosono1,Hosono2,Chen1,Ren1,Ren2,Wen1}, as well as 25 K in hole
doped La$_{1-x}$Sr$_x$OFeAs compound \cite{Wen2}. Because of the
layered structure, the doping behavior and many other properties of
the iron-based system are very similar to those of the
copper-oxides, it has been thus expected that higher $T_c$ may be
found in multi-layer systems. Soon after, single crystals of
LnFeAs(O$_{1-x}$F$_x$) (Ln=Pr, Nd, Sm) were grown successfully by
NaCl/KCl-flux method \cite{Quebe, Zhigadlo, Hashimoto}, while the
sub-millimeter sizes limit the experimental study on them
\cite{Wen3, Weyeneth}. Therefore, FeAs-based single crystals with
high crystalline quality, homogeneity and large sizes are highly
desired for precise measurements of the properties.

Very recently, the BaFe$_2$As$_2$ compound in a tetragonal
ThCr$_2$Si$_2$-type structure with infinite Fe-As layers was
reported \cite{Johrendt1}. By replacing the alkaline earth elements
( Ba and Sr ) with alkali elements ( Na, K and Cs ),
superconductivity up to 38 K was discovered both in hole-doped and
electron doped samples \cite{Johrendt2, Chen2, Wang1, Chen3}. The
$T_c$ varies from 2.7 K in CsFe$_2$As$_2$ to 38 K in
A$_{1-x}$K$_x$Fe$_2$As$_2$ (A=Ba, Sr) \cite{Johrendt2, Chu}.
Meanwhile, the superconductivity also could be induced in the parent
phase by high pressure \cite{Canfield3, Thompson1} or replacing
partial Fe by Co \cite{Jasper, Sefat}.  More excitingly, large
single crystals could be obtained by Sn-flux method in this family
for the rather low melting temperature and the intermetallic
characteristics \cite{Canfield1, Wang2, Thompson2}. However, single
crystals with high homogeneity and less contamination are still hard
to be obtained by this method \cite{Canfield2}. To avoid these
problems, the FeAs self-flux method may be more appropriate.

Here we report the successful growth of A$_{1-x}$K$_x$Fe$_2$As$_2$
(A=Ba,Sr) single crystals by self-flux method using FeAs as the
flux, both non-superconducting parent phase AFe$_2$As$_2$ (A=Ba,Sr)
and superconducting Ba$_{1-x}$K$_x$Fe$_2$As$_2$ crystals are
obtained. The measurements of X-ray diffraction (XRD) indicate high
crystalline quality on both kinds of samples, and the results of AC
susceptibility show sharp superconducting transitions on the
superconducting ones. The resistivity measurements on the
non-superconducting phase suggest a clear resistivity anomaly which
is induced by the formation of the spin-density-wave (SDW) and
structure transition \cite{Johrendt1, Dai, Johrendt4, Chen5}. The
doping dependence of $c$-axis lattice constant and $T_c$ are very
consistent with the reported results from polycrystalline samples
\cite{Chen6, Johrendt3}.

\section{Experiments}

The A$_{1-x}$K$_x$Fe$_2$As$_2$ (A=Ba,Sr) single crystals were grown
by FeAs self-flux method.  The FeAs precursor was synthesized by the
reaction of Fe powders (Alfa Aesar, 99.99\% in purity) and As chips
(99.999\%) at 500 $^o$C for 10 hours and then 700 $^o$C for 20 hours
in a sealed silica tube. The starting materials of FeAs, and high
purity Ba or Sr (Alfa Aesar, 99.2\% in purity) were mixed in
4:(1-$x$), then a soft bulk of K with proper amount was added to
cover the powder. The whole procedure was performed in a glove box
with a protective argon atmosphere where both concentrations of
O$_2$ and H$_2$O were less than 1 ppm. The mixture was placed in an
alumina oxide crucible with $\phi$15 mm $\times$ 35 mm in dimension
and sealed under vacuum in a silica tube with $\phi$18 mm $\times$
90 mm in dimension. Because the silica tube would be broken due to
the gas pressure of potassium at the temperature around 1000
$^\circ$C, the superconducting samples could only be obtained by
using an limited amount of potassium and a thick enough silica tube
( t $>$ 3 mm ). For example, if the total mass of staring material
was supposed to be 2.0 g with the ratio of Ba : K : Fe : As = 0.6 :
0.4 : 4.0 : 4.0, the mass for each materials was m(Ba)= 0.265 g,
m(K)= 0.050 g and m(FeAs)= 1.684 g, respectively. Considering the
losing of K during the growth (about 0.12g in most cases), the total
amount of K should be 0.170 g. It should be noted that the safe
amount of K is less than 0.25g in our conditions for preventing the
explosion of silica tube. The actual contents of K in the as-grown
crystals were determined by quantitative analysis in the later
measurements. The sealed silica tube was placed in a muffle furnace
and heated up to a high temperature at about 1000 $\sim$ 1150 $^o$C
for melting completely. Then it was cooled down to a temperature
below 800 $^o$C at a very slow speed which is less than 10
$^o$C/hour. The melting temperature and cooling down speed depend on
the ratio of Ba:K in the staring material. Finally, the muffle
furnace was powered off. After it was cooled down to room
temperature, the tube was fetched out and broken. The crystals were
obtained by cleaving the as-grown bulks. Then they were selected and
shaped under a microscope.

Various techniques were used to characterize our samples. The
crystal surface morphology and composition were examined by scanning
electron microscopy (SEM, Hitachi S-4200) and the energy dispersive
X-ray (EDX, Oxford-6566, installed in the S-4200 apparatus)
analysis. While the X-ray diffraction of the crystals was carried
out by a \emph{Mac- Science} MXP18A-HF equipment with
$\theta-2\theta$ scan to examine the crystalline quality of the
samples. $K_{\alpha}$ radiation of Cu target was used, and the
continuous scanning range of 2$\theta$ is from 10$^{o}$ to 80$^{o}$.
The raw data of XRD was analyzed by \emph{PowderX} software where
the zero-shift and $K_{\alpha2}$-elimination and other factors were
taken into account \cite{Dong}. The AC susceptibility was measured
on an \emph{Oxford }cryogenic system Maglab-EXA-12. An alternating
magnetic field (H=1 Oe) was applied perpendicular to the $ab$-plane
at a frequency $f$=333 Hz when the AC susceptibility measurement was
undertaken. The $T_c$ was derived from AC susceptibility curve by
the point where the real part of the susceptibility becomes flat.
The resistivity measurements were carried out on a \emph{Quantum
Design} Physical Property Measurement System (PPMS) by a standard
four-probe method with a low contact resistance ($<1$ $\Omega$).

\section{Results and discussion}

By varying the content of potassium in the starting material, we
obtained non-superconducting crystals in both Ba-based and Sr-based
compound and superconducting samples with the composition of
Ba$_{1-x}$K$_x$Fe$_2$As$_2$. Figure 1 shows the photograph of some
crystals cleaved from the as-grown bulks. They all have very shiny
plate-like cleaved surfaces. The sizes of the largest one are about
2.5 mm$\times$4.0 mm$\times$0.2 mm, and others have sizes up to 2
mm. Because the real contents of each element always deviate from
the starting material in the flux method. The composition of our
single crystals was determined by the energy dispersive X-ray (EDX)
analysis. 3 $\sim$ 5 pieces of as-grown single crystal were selected
out carefully from each batch. Then they were cleaved under
microscope and taken EDX measurement immediately before the surface
degenerate in air. A typical EDX spectrum is shown in Figure 2. The
inset is the SEM photograph of this crystal, which shows a very flat
surface morphology and the layered structure.  A brief summary of
the properties of Ba$_{1-x}$K$_x$Fe$_2$As$_2$ single crystals is
given in Table 1.  We successfully obtained four categories
superconducting samples with K doping level in $x$= 0.40, 0.28, 0.25
and 0.23.  The non-superconducting crystals also contain a little K
which is less than 10\% for both Ba-based and Sr-based compound.

The crystal structure of non-superconducting samples was examined by
X-ray diffraction measurement with incident X-ray along the
$c$-axis. The typical diffraction patterns are shown in Figure 3.
Only sharp peaks along ($00l$) could be observed, and
full-width-at-half-maximum (FWHM) of each peak is around
0.10$^{\circ}$.  These indicate high $c$-axis orientation and
crystalline quality in our samples. The raw data of XRD was analyzed
by \emph{PowderX} software with the zero-shift and
$K_{\alpha2}$-elimination and other factors taken into account
\cite{Dong}.  The $c$-axis parameters were calculated and also
presented in Table 1.  For the non-superconducting samples, the
$c$-aixs lattice constant is about 13.07 \AA\ for Ba-based compound
and 12.58 \AA\ for Sr-based compound, respectively.  The magnitudes
of $c$-axis are very close to those results in the polycrystalline
parent phase without K doing, where they were reported to be 13.02
\AA\ \cite{Johrendt1} for Ba-based compound and 12.40 \AA\
\cite{Wang1} for Sr-based compound.  In addition, it should be noted
that the EDX has a uncertainty around 10\%, and it will be larger
especially for light elements. It is possible that the real contents
of K in non-superconducting samples are less than the magnitude
shown in Table 1. Thus we just mark them using the chemical formula
as parent compounds BaFe$_2$As$_2$ and SrFe$_2$As$_2$.  Figure 3(b)
also displays the XRD pattern for the sample with $x$= 0.40. A clear
shift for each peak shows up, which indicates that the lattice has a
little variation after doping K into the parent phase.

The AC susceptibility measurement was used to character the
superconducting Ba$_{1-x}$K$_x$Fe$_2$As$_2$ single crystals.  Figure
4 shows three typical groups of the susceptibility curves.  The
$T_c$ was determined as the onset point of $\chi'$, and the
transition width was defined as $\Delta T_c$=$T_c$(90\%) -
$T_c$(10\%). There are a flat diamagnetic signal in low temperature
region and a very sharp superconducting transition around $T_c$,
where the demagnetizing factor is not taken into account in these
measurements. The $T_c$ increases gradually as more and more K were
doped into the samples.  The superconducting transition is almost
the same among the single crystals cleaved from the same batch,
which indicates that our samples are very homogeneous.

Figure 5 shows the temperature dependence of resistivity under zero
field.  The applied current is 5 mA, and it flows in the $ab$-plane
during the measurements. For BaFe$_2$As$_2$ and SrFe$_2$As$_2$, a
strong anomaly shows up at $T_s$= 135 K and 173 K, respectively (
Figure 5(a) ). The resistivity has a nearly $T$-linear dependence
above this temperature and sharply drops down below this
temperature. This resistivity anomaly could be attributed to the SDW
instability and structure transition which was also observed in
other systems \cite{Wang1, Dai, Johrendt4, Chen5}. However, the
characteristic temperatures found in our samples are lower than
those in other reports both for BaFe$_2$As$_2$ ($T_s$= 140 K)
\cite{Johrendt1, Chen5} and SrFe$_2$As$_2$ ($T_s$= 195 K)
\cite{Wang1}. It may be induced by doping a little bit of potassium
into the parent compounds. Increasing hole-doping further will
suppress the SDW transition, and superconductivity eventually
emerges \cite{Chen6}. Therefore, the superconducting
Ba$_{1-x}$K$_x$Fe$_2$As$_2$ samples were obtained by adding more K
into the starting material. Figure 5(b) shows the temperature
dependence of resistivity for the superconducting single crystals.
The SDW anomaly is smeared in the normal state, and a
superconducting transition emerges at lower temperatures.  The
$T_c$(conset)s for different doping are about 36.6 K, 31.4 K, 28.7 K
and 24.5 K. The resistivity data also indicates sharp transition in
our samples with $\Delta T_c$= 0.44, 0.49, 0.71 and 0.40 K ( 90\% -
10\% of normal state resistivity ). Furthermore, if we extrapolate
the data just above the superconducting transition by a straight
line, it could be roughly estimated that the residual resistivity is
almost close to zero for the sample with $T_c$= 36 K. This indicates
that our samples are rather clean.

When preparing this paper, we became aware that two papers working
on the Ba$_{1-x}$K$_x$Fe$_2$As$_2$ polycrystalline samples with
series doping levels were posted on arXiv \cite{Chen6, Johrendt3}.
Thus we made a comparison on the doping dependence of $c$-axis and
$T_c$ between the polycrystalline samples and our single crystals.
The result is shown in Figure 6.  Our data is very consistent with
the data in Reference [31], while a little deviation is found for
the data in Reference [30]. However, the general behaviors are
almost the same between the polycrystalline samples and single
crystals. The $c$-axis expands almost linearly as increasing the
content of K. While the $T_c$ increases quickly as a little K was
doped into the parent compound, then it grows slowly between $x$=
0.3 $\sim$ 0.4. It seems that the maximal of $T_c$ is achieved
between $x$= 0.4 $\sim$ 0.5. Therefore, our samples with $x$= 0
$\sim$ 0.4 reside in the "underdoped" regime. Growth of the
"overdoped" single crystals is underway.

\section{Summary}

In summary, we have successfully grown the single crystals of
A$_{1-x}$K$_x$Fe$_2$As$_2$ (A=Ba, Sr) with high quality by using
FeAs as the self-flux.  By varying the K content during the growth,
we obtained non-superconducting Ba(Sr)Fe$_2$As$_2$ single crystals
and superconducting Ba$_{1-x}$K$_x$Fe$_2$As$_2$ single crystals with
$x$= 0.23, 0.25, 0.28 and 0.40.  The samples have sizes up to 4 mm
with flat and shiny cleaved surfaces. The X-ray diffraction patterns
with only (00$l$) peaks suggest that they have high crystalline
quality. The superconductivity is characterized by AC susceptibility
and resistivity measurements which exhibit very sharp
superconducting transitions. While the temperature dependence of
resistivity on the non-superconducting crystals shows that the SDW
instability and structure transition occur at about 173 K and 135 K
for Sr-based and Ba-based compound, respectively.  The doping
dependence of $c$-axis parameters and $T_c$ are consistent with the
previous data from polycrystalline samples \cite{Chen6, Johrendt3},
which indicates the effects of different potassium doping.

\textbf{Acknowledgement}

This work was financially supported by the Natural Science
Foundation of China, the Ministry of Science and Technology of China
(973 Projects Nos. 2006CB601000, 2006CB921802 and 2006CB921300), and
Chinese Academy ofSciences (Project ITSNEM). The authors acknowledge
the helps from Lihong Yang and Hong Chen for the XRD measurements,
and the helpful discussions with Lei Fang, Lei Shan and Cong Ren at
IOP, CAS.

\newpage

\begin{table}
\caption{\label{tabone}A brief summary of the properties of
Ba$_{1-x}$K$_x$Fe$_2$As$_2$ single crystals: The actual cationic
compositions (Ba:K) determined by EDX, $c$-axis parameters, critical
temperature, and typical sizes of crystals.}
\begin{indented}
\lineup
\item[]\begin{tabular}{@{}*{5}{l}}
\br

        \0\0 no.  &\0Ba\0:\0K   & $c$ (\AA)   & $T_c$(K)   & typical sizes
(mm$^3$)\cr\mr
        \0\m 1   &0.60\0:\00.40  & 13.344  &36.3 &$1\times1\times0.1$    \cr
        \0\m 2   &0.72\0:\00.28  & 13.226  &31.8 &$4\times3\times0.2$    \cr
        \0\m 3   &0.75\0:\00.25  & 13.196  &27.5 &$3\times2\times0.1$    \cr
        \0\m 4   &0.77\0:\00.23  & 13.189  &23.4 &$2\times1\times0.1$    \cr
        \0\m 5   &0.92\0:\00.08  & 13.077  &\00    &$1\times0.5\times0.1$  \cr
        \0\m 6   &0.94\0:\00.06  & 13.065  &\00    &$3\times2\times0.3$    \cr \br
\end{tabular}
\end{indented}
\end{table}

\begin{figure}
  \center\includegraphics[width=2.8in]{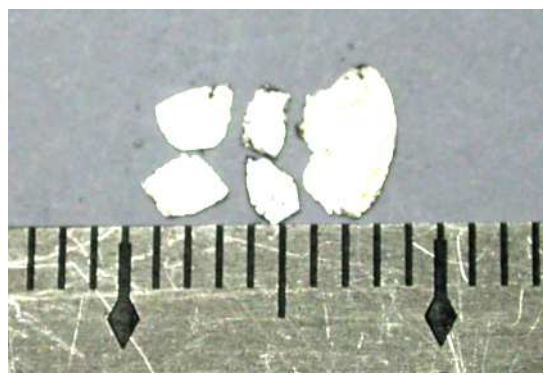}\\
  \caption{Photograph of the A$_{1-x}$K$_x$Fe$_2$As$_2$
(A = Ba,Sr) crystals cleaved from the as-grown bulk. The crystals
have rather shiny surfaces with sizes up to 4 mm. }
  \label{f1}
\end{figure}

\begin{figure}
   \center\includegraphics[width=3.6in]{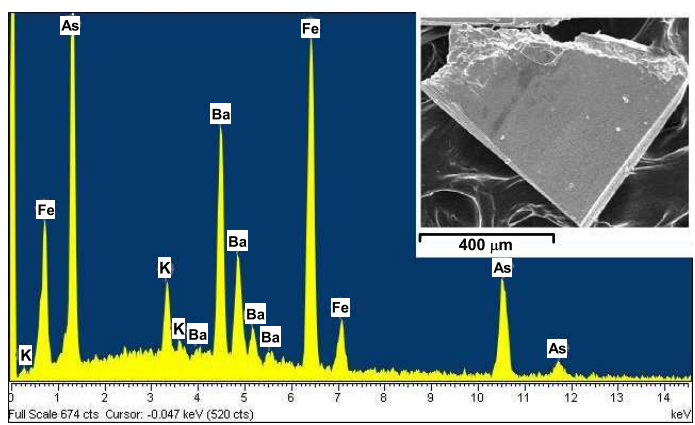}\\
  \caption{A typical EDX spectrum for one single crystal.
  The inset is the SEM photograph of this crystal, which shows a very flat surface morphology and the layered structure.}
  \label{f2}
\end{figure}

\begin{figure}
   \center\includegraphics[scale=1]{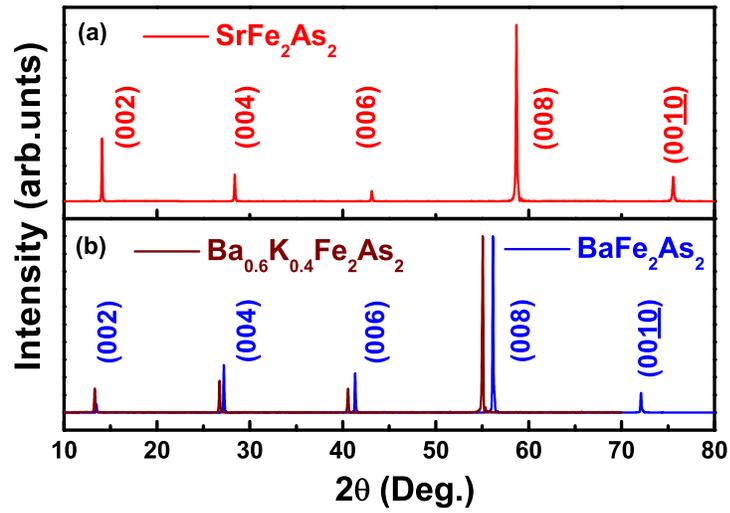}\\
  \caption{Typical XRD patterns for cleaved crystals. The FWHM of each peak is around
0.10$^{\circ}$. A clear shift was observed after doping 40\% amount
of K in the Ba$_{1-x}$K$_x$Fe$_2$As$_2$ system. }
  \label{f3}
\end{figure}

\begin{figure}
  \center\includegraphics[width=3.2in]{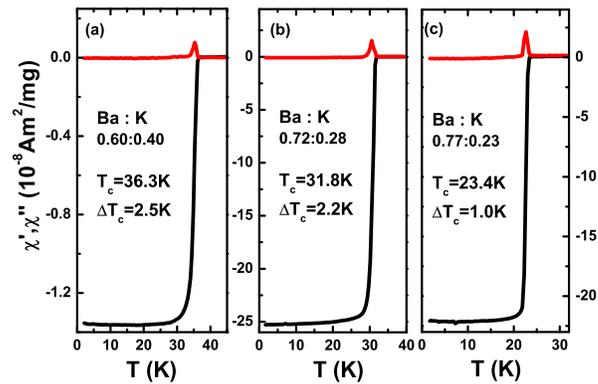}\\
  \caption{Temperature dependence of AC susceptibility for superconducting Ba$_{1-x}$K$_x$Fe$_2$As$_2$ single crystals. The $T_c$ was derived by the point
where the real part of the susceptibility becomes flat, and the
transition width was defined as $\Delta T_c$=$T_c$(90\%) -
$T_c$(10\%).}
  \label{f4}
\end{figure}

\begin{figure}
  \center\includegraphics[width=3.0in]{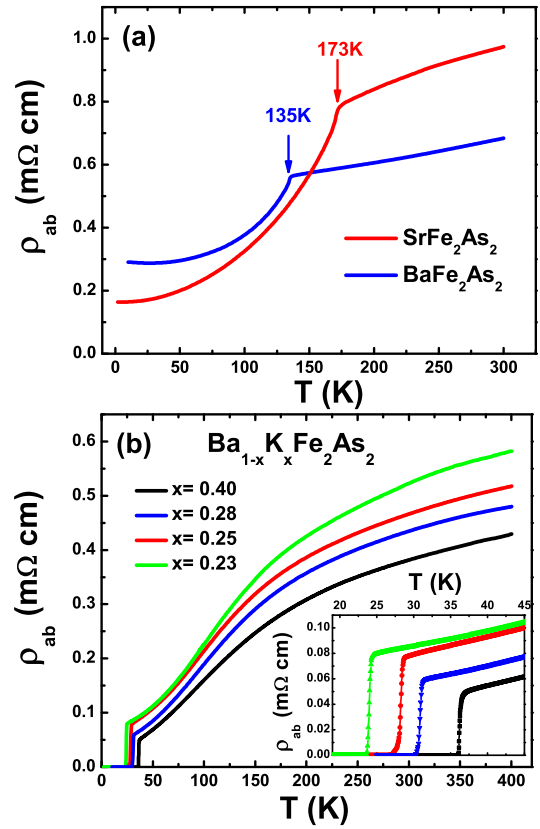}\\
  \caption{(a).Temperature
dependence of resistivity for BaFe$_2$As$_2$ and SrFe$_2$As$_2$
crystals. The SDW anomaly happens at 135 K and 173 K, respectively.
(b).Temperature dependence of resistivity for the superconducting
single crystals. The inset is the zooming in graph around the
superconducting transition.}\label{f5}
\end{figure}

\begin{figure}
  \center\includegraphics[width=3.0in]{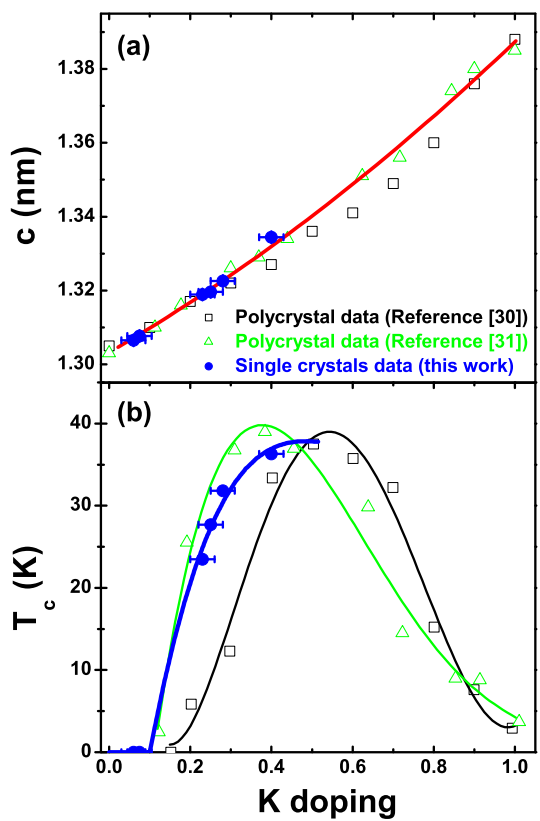}\\
  \caption{ Doping dependence of the $c$-axis and $T_c$ for our single crystals (blue points). The open black squares and green
  triangles are polycrystalline data from ref. 30 and ref. 31, respectively.
   }\label{f6}
\end{figure}
\end{document}